\documentclass[a4paper,USenglish]{lipics-v2016}

\usepackage[tikz]{bclogo}
\usepackage{url}
\usepackage{amssymb}
\usepackage{graphicx}
\usepackage{amsmath}
\usepackage{amsfonts}
\usepackage{color}

\title{Entanglement Verification in Quantum Networks with Tampered Nodes}
\titlerunning{Entanglement Verification in Quantum Networks with Tampered Nodes}

\author{Michele Amoretti$^1$, Stefano Carretta$^2$}
\affil{1: Department of Engineering and Architecture - University of Parma, Italy} 
\affil{2: Department of Mathematical, Physical and Computer Sciences - University of Parma, Italy} 
\affil{Quantum Information Science - University of Parma, Italy}
\affil{Contact: michele.amoretti@unipr.it}
\authorrunning{M. Amoretti, S. Carretta} 


\begin{document}

\maketitle

\begin{abstract}
In this paper, we consider the problem of entanglement verification across the quantum memories of any two nodes of a quantum network. Its solution can be a means for detecting (albeit not preventing) the presence of intruders that have taken full control of a node, either to make a denial-of-service attack or to reprogram the node. Looking for strategies that only require local operations and classical communication (LOCC), we propose two entanglement verification protocols characterized by increasing robustness and efficiency.
\end{abstract}

\section{Introduction}

Quantum networking is an emerging field using the properties of quantum mechanics to bring new, useful capabilities to networking. Quantum networks enable the transmission of quantum information between physically separated quantum computers \cite{VanMeter2014,Gyongyosi2018a,Gyongyosi2018d}. In particular, they support the end-to-end generation of \textit{entangled quantum states} across distant nodes \cite{Gyongyosi2019a,Gyongyosi2018c,Gyongyosi2018e,Gyongyosi2019b}. Entangled states exhibit correlations that have no classical analog and may be used, e.g., to solve leader election problems \cite{Tani2012}, to perform distributed computing tasks \cite{Buhrman2003,Amoretti2019}, to share secrets \cite{BB84,Ekert1991,Vazirani2014}, or to perform remote synchronization of clocks \cite{BenOr2005}. The IRTF Quantum Internet Research Group (QIRG), whose purpose is to standardize quantum networking protocols, has recently released an Internet-Draft on advertising entanglement capabilities \cite{QIRG}.

In this work, we consider the problem of entanglement verification across the quantum memories of any two nodes of a quantum network. Its solution can be a means for detecting (albeit not preventing) the presence of intruders that have taken full control of a node. We focus on \textit{Bell states}, i.e., maximally entangled states of quantum systems made of two qubits. In particular, we consider two-party Bell pairs, where the qubits of each pair are physically separated, stored in the quantum memories of two distant nodes of the quantum network.

Experimental procedures for entanglement verification have been classified by van Enk \textit{et al.} \cite{vanEnk2007} as follows: 1) teleportation, 2) Bell-CHSH inequality tests, 3) tomography, 4) entanglement witnesses, 5) direct measurement of entanglement, 6) consistency with entanglement.
All these approaches have been studied analytically and experimentally, in the context of scenarios that are different from the one described in this paper --- where the source of entanglement is assumed to be trusted, but one party may be dishonest. Usually, the focus is either on untrusted sources of entangle qubits (photons, in most cases) or on eavesdroppers that interfere with the quantum channel while entangled qubits are transmitted. 
 
Blume-Kohout \textit{et al.} \cite{BlumeKohout2010} illustrated a reliable method to quantify exactly what can be concluded from finite data sets resulting from measurements in entanglement verification protocols. The authors made no assumption on the causes of non-entanglement. 
Christandl and Renner \cite{Christandl2012} showed that quantum state tomography, together with an appropriate data analysis procedure, allows one to obtain confidence regions, i.e., subsets of the state space in which the true state lies with high probability. The proposed approach can be applied to arbitrary measurements including fully coherent ones, as shown by Arrazola \textit{et al.} \cite{Arrazola2013}. 

Most protocols refer to the scenario in which entangled qubits are photon pairs, as it is the most simple to implement experimentally. Usually, there are two parties that do not trust each other and may (or may not) trust the source of entangled photon pairs.
Bennett \textit{et al.} \cite{Bennett2012} proposed a protocol that allows one party to verify entanglement when the (untrusted) Prover is also the source of entangled photon pairs.
Moroder \textit{et al.} \cite{Moroder2013} presented a framework for device-independent quantification of bi- and multipartite entanglement, meaning that 
the amount of entanglement is measured based on the observed classical data only but independent of any quantum description of the employed devices. In their problem formulation, the authors do not consider the possibility that involved parties may be dishonest.

We look for entanglement verification protocols that only require local operations and classical communication (LOCC) to work, which are easier to implement and more robust to channel noise. The LOCC protocol proposed by Nagy and Akl \cite{Nagy2010b} (denoted as \textsf{NA2010} for simplicity) is a working solution that we consider as a baseline. 

Two-party LOCC protocols for entanglement verification can be compared in terms of robustness and efficiency. The general definition of robustness suggested by Renner \cite{Renner2005} is a perfect fit.
\begin{definition} 
A two-party protocol $\mathcal{P}$ is \textit{$\epsilon$-robust} on the inputs of the two interacting parties if the probability that the protocol aborts is at most $\epsilon$. 
\end{definition}
For the efficiency, we adopt a more specific definition.
\begin{definition} 
A two-party entanglement verification protocol $\mathcal{P}_1$ is more efficient than another two-party entanglement verification protocol $\mathcal{P}_2$, if the probability that the protocols abort is the same, but $\mathcal{P}_1$ sacrifices less Bell states than $\mathcal{P}_2$.
\end{definition}

\subsection{Contributions}
We state the entanglement verification problem according to the single-prover quantum interactive proof model \cite{Vidick2016}. The two parties are denoted as Verifier and Prover, respectively. The Verifier wants to check the honesty of the Prover, which is untrusted.

As a warm up, we summarize the reference protocol \textsf{NA2010} \cite{Nagy2010b} and we formalize its analysis by proving the following theorem.

\begin{theorem}  
\textsf{NA2010} is $(7/8)^m$-robust on any set of $m$ Bell states shared by the Verifier and the Prover, assuming that the Prover is an attacker that performs measurements either in the computational or diagonal basis.
\end{theorem}

Then, we propose two entanglement verification protocols, denoted as \textsf{AC1} and \textsf{AC2}, which are characterized by increasing robustness and efficiency. 
More specifically, we prove the following theorems. 

\begin{theorem} 
\textsf{AC1} is $(3/4)^m$-robust on any set of $m$ Bell states shared by the Verifier and the Prover, assuming that the Prover is controlled by an attacker that performs measurements either in the computational or diagonal basis.
\end{theorem}

\begin{theorem}  
\textsf{AC2} is $(3/8)^m$-robust on any set of $2m$ Bell states shared by the Verifier and the Prover, assuming that the Prover is controlled by an attacker that performs measurements either in the computational or diagonal basis.
\end{theorem}

We also show that, remarkably, the success probability of each \textsf{AC2} round is always $\geq 1/2$ \textit{for any measurement basis} the malicious Prover adopts to destroy the entanglement. Noteworthy, this result holds despite the assumption that the attacker has exactly the same capabilities of the Verifier. Moreover, we characterize and compare the three protocols in terms of efficiency, showing that \textsf{AC2} is better than \textsf{AC1}, which is better than \textsf{NA2010}. 

Last but not least, we illustrate simulation results, obtained with SimulaQron \cite{Dahlberg2017}, and experimental results, based on the IBM Q platform \cite{IBMQ}, that confirm the theoretical analysis of the three protocols.

\section{Notation}
\label{sec:Notation}
The notation adopted in this paper is the usual one in quantum computing \cite{NC2000,Gyongyosi2018b}.
A qubit is a quantum-mechanical system whose state can be represented as a vector in $\mathbb{C}^2$. Using the \textit{computational basis} $|0\rangle = 
\begin{pmatrix}
1 \\
0
\end{pmatrix}, |1\rangle = 
\begin{pmatrix}
0 \\
1
\end{pmatrix}$, 
the generic state of a qubit is $|\psi\rangle = \alpha |0\rangle + \beta |1\rangle$, where $\alpha, \beta \in \mathbb{C}$ are called \textit{probability amplitudes} (with $|\alpha|^2 + |\beta|^2 = 1$). 

When a qubit gets measured, the result is either $0$ or $1$, with probability $|\alpha|^2$ or $|\beta|^2$, respectively. 
The post-measurement state of the qubit is either $|0\rangle$ or $|1\rangle$, respectively. 
Thus, if a qubit results to be in superposition of the basis states, with unknown probability amplitudes, there is no way to know the values of $\alpha$ and $\beta$ with a single measurement. 

One may use \textit{diagonal basis}
$|+\rangle = \frac{|0\rangle + |1\rangle}{\sqrt{2}}, 
|-\rangle = \frac{|0\rangle - |1\rangle}{\sqrt{2}}$
to represent the state of a qubit: 
$|\psi\rangle = \frac{\alpha + \beta}{\sqrt{2}}|+\rangle + \frac{\alpha - \beta}{\sqrt{2}}|-\rangle$.
Measuring with respect to the diagonal basis results in $+$, with probability $|\alpha + \beta|^2/2$, or $-$, with probability $|\alpha - \beta|^2/2$. The corresponding post-measurement states are $|+\rangle$ or $|-\rangle$, respectively.

The Hadamard operator $H = \frac{1}{\sqrt{2}} \begin{pmatrix} 1 & 1 \\ 1 & -1\end{pmatrix}$ maps $|0\rangle$ to $|+\rangle$ and $|1\rangle$ to $|-\rangle$. Other operators we will use are Pauli-X: 
$X = \begin{pmatrix} 0 & 1 \\ 1 & 0\end{pmatrix}$, and Pauli-Z:
$Z = \begin{pmatrix} 1 & 0 \\ 0 & -1\end{pmatrix}$.

An $n$ qubit system has $2^n$ basis states.
Any measurement result $x \in \{0,1\}^n$ occurs with probability $|\alpha_x|^2$, $\alpha_x \in \mathbb{C}$, with the state of the qubits after the measurement being $|x\rangle$. The normalization condition is $\sum_{x \in \{0,1\}^n} |\alpha_x|^2 = 1$.

Sometimes it is not possible to decompose the state of an $n$ qubit quantum system in the tensor product of the component states. Such a state is denoted as \textit{entangled}. Bell states are maximally entangled quantum states of two qubits:
$|\beta_{00}\rangle = \frac{ |00\rangle + |11\rangle }{\sqrt{2}}, 
|\beta_{01}\rangle = \frac{ |01\rangle + |10\rangle }{\sqrt{2}},
|\beta_{10}\rangle = \frac{ |00\rangle - |11\rangle }{\sqrt{2}},
|\beta_{11}\rangle = \frac{ |01\rangle - |10\rangle }{\sqrt{2}}$.
If we measure the first qubit, in $|\beta_{00}\rangle$ we obtain either $0$ or $1$. Result $0$ leaves the post-measurement state $|00\rangle$, while result $1$ leaves $|11\rangle$. Thus, a measurement of the second qubit always gives the same result as the measurement of the first qubit. The same reasoning applies to the other types of Bell states.

\section{Attack Model}   
\label{sec:AttackModel}

In quantum networks, the following attack scenarios may be considered.
\begin{enumerate}
\item The attacker is able to measure some (or all) of the qubits---either by intercepting them while they travel on a network link, or by gaining access to the memory of some nodes.
\item The attacker is able to entangle some (or all) of the qubits with a quantum memory of her own, and measure that memory at a later point after listening in on the classical communications.
\item The attacker is able to completely take over certain nodes, including the quantum memories at those nodes. This is essentially the assumption that some parties are not trustworthy.
\end{enumerate}
It is assumed that the attacker can listen in on all classical communications.

Work on \textit{quantum key distribution} has generally shown that scenario 2 does not give the attacker appreciably more power than scenario 1. By contrast, scenario 3 makes the attacker much more powerful. To detect the attacker while intercepting qubits that are transmitted over quantum channels, there are several known effective strategies. For example, the generalized Bell's theorem (CHSH inequalities) \cite{CHSH1969} can be used, as suggested by Ekert \cite{Ekert1991}.

In this paper we focus on scenario 3, where the attacker physically captures the node and takes full control of its functioning. Such a node is generally called \textit{compromised}, but there are two alternative cases.
In the first case, the attacker interacts with the local quantum memory, reconfiguring the states of the qubits (e.g., breaking entangled states shared with other nodes, by measuring local qubits) either to make a denial-of-service attack or to reprogram the node to a behavior in accordance with her own plans. Thus, entanglement verification is a means for intrusion detection. In the second case, which is out of the scope of this paper, the attacker does not interact with the local quantum memory, thus preserving the entangled quantum states that are shared with other nodes.
Thus, the attacker can read incoming messages and send her own ones, but she cannot reprogram the node. The success of any verification scheme relies on the attacker's weakness in making the node behave as if it was not compromised. 

Definitely, the attack scenario we consider is very disruptive. However, the entanglement verification protocols we propose put a limit on the actions the all-powerful attacker may do. Indeed, there is no perfect strategy for the attacker, due to the design of the protocols.

\section{Entanglement Verification Protocols}   
\label{sec:EntVerif}

In the following, we propose a formal analysis of the \textsf{NA2010} LOCC protocol for entanglement verification proposed by Nagy and Akl \cite{Nagy2010b}, which is the one we consider as a baseline.
Then, we define and analyze two alternative LOCC protocols for entanglement verification, namely \textsf{AC1} and \textsf{AC2}. Compared to \textsf{NA2010} \cite{Nagy2010b}, \textsf{AC1} is much more robust and efficient. \textsf{AC2} is even better than \textsf{AC1}.

\subsection{\textsf{NA2010} Protocol} 
\begin{figure}[h!]
\centering
\includegraphics[width=5.5cm]{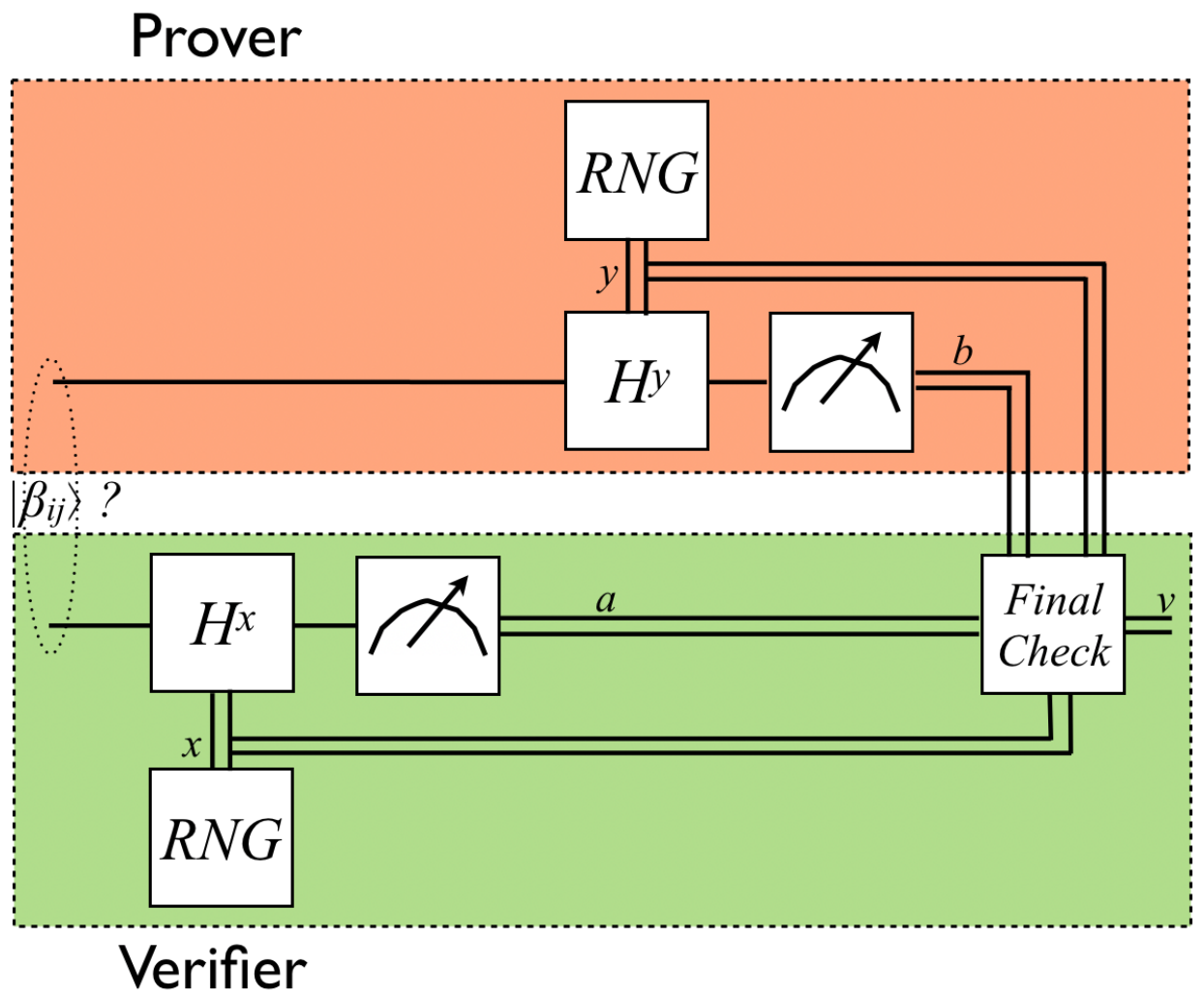}
\caption{Quantum circuit for \textsf{NA2010}.}
\label{fig:NA2010}
\end{figure}

Let us suppose that two nodes share $n$ entangled qubit pairs whose assumed state is $|\beta_{ij}\rangle$, with $i,j \in \{0,1\}$. One of the two nodes, denoted as the Verifier, wants to check if the other node (the Prover) has been compromised. To this purpose, the Verifier starts the \textsf{NA2010} protocol illustrated in Fig. \ref{fig:NA2010} and detailed in Box \ref{box:NA2010}.

\begin{bclogo}[arrondi=0.2, logo=\bccube]{ Box \ref{box:NA2010} - \textsf{NA2010} Protocol}	
\label{box:NA2010}
\begin{scriptsize}
\noindent
$v \leftarrow 0$, $k \leftarrow 0$ \quad where $v \in \{0,1\}, k \in \mathbb{N}$\\
WHILE ($(k<m) \wedge (v=0)$) DO
\begin{enumerate}
\item $k \leftarrow k+1$
\item The Verifier picks at random a previously unchecked qubit pair (with assumed state $|\beta_{ij}\rangle$) and a random value $x \in \{0,1\}$.
\item If $x=0$, the Verifier measures in the computational basis its qubit of the selected qubit pair. If $x=1$, the Verifier applies the Hadamard operator $H$ to its qubit of the selected qubit pair, then measures the qubit in the computational basis (this is equivalent to measuring in the diagonal basis). The measured value is denoted as $a$.
\item Using a public and authenticated classical channel, the Verifier sends a message to the Prover, specifying which qubit pair is being checked (by means of an identifier).
\item The Prover selects a random value $y \in \{0,1\}$. 
\item If $y=0$, the Prover measures its qubit of the selected qubit pair. If $y=1$, the Prover applies the Hadamard operator to its qubit of the selected qubit pair, then measures the qubit.
\item Using a public and authenticated classical channel, the Prover sends the measured value $b$ and $y$ to the Verifier.
\item The Verifier produces a classical bit $v=1$ if one of the following conditions hold:
\begin{enumerate}
\item $|\beta_{00}\rangle$ or $|\beta_{10}\rangle$, $x=y=0$, $a \neq b$;
\item $|\beta_{10}\rangle$ or $|\beta_{11}\rangle$, $x=y=0$, $a = b$;
\item $|\beta_{00}\rangle$ or $|\beta_{01}\rangle$, $x=y=1$, $a \neq b$;
\item $|\beta_{10}\rangle$ or $|\beta_{11}\rangle$, $x=y=1$, $a = b$;
\end{enumerate} 
In any other case, the Verifier produces $v=0$. 
\end{enumerate}  
IF $v=1$ THEN output ``compromised'' ELSE abort.  
\end{scriptsize}
\end{bclogo}

\subsubsection{Robustness Analysis of \textsf{NA2010}}  

\begin{proof}[Proof of Theorem 1]  
If the Prover has not been compromised, its behavior is expected to be fair, i.e., respecting the protocol. If the Prover is an attacker, it could, in theory, send random $b$ and $y$ values to the Verifier. However, this would not affect the chances of being detected by the Verifier (as the Prover does not know $x$). Thus, we assume the Prover respects the protocol also when it is an attacker.

Consider the case where the assumed state of the selected qubit pair is $|\beta_{00}\rangle$.
If the Prover is controlled by an attacker that has measured its qubit in the computational basis, the actual state is either $|00\rangle$ or $|11\rangle$. If $x=y=0$, measuring returns either $00$ or $11$, with equal probability. Thus, it is not possible to detect that the Prover is an attacker.
If $x=y=1$ (whose probability is $1/4$), measurement results may be $00$, $01$, $10$ or $11$, with equal probability. When $a \neq b$ (i.e., measurement results are either $01$ or $10$; the probability is $1/2$), the attacker is detected. 
If the Prover is controlled by an attacker that has measured its qubit in the diagonal basis, the actual state is either $|++\rangle$ or $|--\rangle$. If $x=y=0$, measurement results may be $00$, $01$, $10$ or $11$, with equal probability. When $a \neq b$, measurement results are either $01$ or $10$), the attacker is detected. Instead, if $x=y=1$, measuring returns either $00$ or $11$, with equal probability. Thus, it is not possible to detect that the Prover has been compromised.
To summarize, when the selected qubit pair is $|\beta_{00}\rangle$, if the attacker measured either in the computational basis or in the diagonal basis, there is always a $1/8$ probability to detect it.
The same result holds for the other Bell states.

Overall, for each round of the protocol, if the attacker performs measurements either in the computational or diagonal basis, the probability to detect a non-entangled pair is $1/8$, i.e., the attacker succeeds with probability $7/8$. If the procedure is executed $m$ times, the protocol aborts with probability $(7/8)^m$.
\end{proof}

In other words, \textsf{NA2010} succeeds in detecting the attacker with probability $p_m = 1 - (7/8)^m$.

\subsubsection{Efficiency Analysis of \textsf{NA2010}} 

\textsf{NA2010} consists of $m$ repetitions of the following operations:
\begin{itemize}
\item random number generation (twice);
\item qubit measurement (twice);
\item application of the $H$ gate (never, once or twice);
\item classical bit dispatching ($k$ by the Verifier, where $k$ is the size of the identifier that specifies which qubit pair is being checked; $2$ by the Prover);
\item binary variable check ($5$ by the Verifier; $k+1$ by the Prover).
\end{itemize}
Only basic single-qubit quantum gates are used. The circuit is specific for \textsf{NA2010}.

By checking $m=10$ qubit pairs with \textsf{NA2010}, the attacker is caught with probability $p_m = 1 - (7/8)^{10} = 0.73$. With $m=20$, the probability is $p_m = 0.93$. To get $p_m = 0.99$, it is necessary to check $m=35$ qubits.

\subsection{\textsf{AC1} Protocol} 
\begin{figure}[h!]
\centering
\includegraphics[width=6cm]{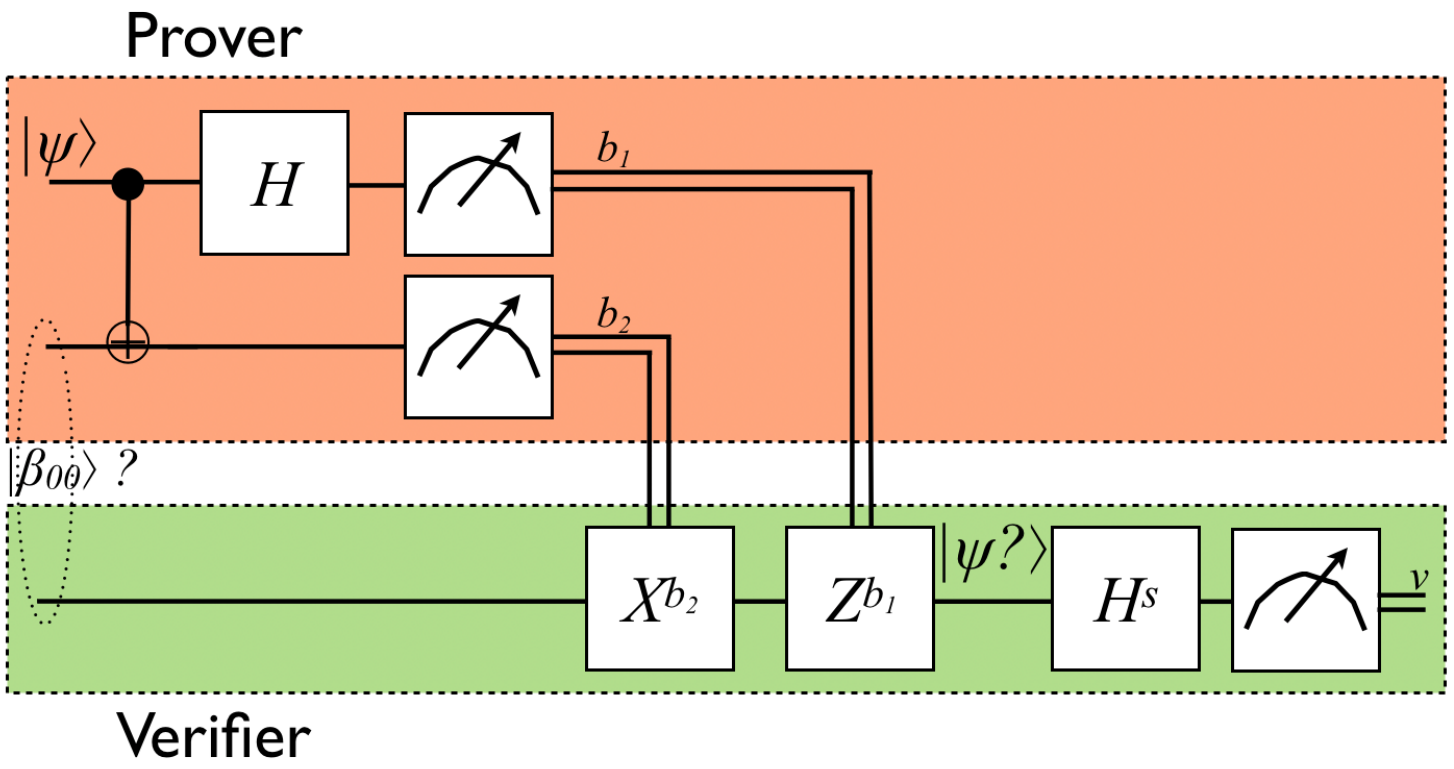}
\caption{Quantum circuit for \textsf{AC1} (case $|\beta_{ij}\rangle = |\beta_{00}\rangle$).}
\label{fig:AC1}
\end{figure}

To check if the Prover has been compromised, the Verifier may start the \textsf{AC1} protocol illustrated in Fig. \ref{fig:AC1} and detailed in Box \ref{box:AC1}.

\begin{bclogo}[arrondi=0.2, logo=\bccube]{ Box \ref{box:AC1} - \textsf{AC1} Protocol}
\label{box:AC1}
\begin{scriptsize}
\noindent
$v \leftarrow 0$, $k \leftarrow 0$ \quad where $v \in \{0,1\}, k \in \mathbb{N}$\\
WHILE ($(k<m)\wedge(v=0)$) DO
\begin{enumerate}
\item $k \leftarrow k+1$
\item Using a public and authenticated classical channel, the Verifier sends a classical message to the Prover, to start a quantum teleportation process \cite{Bennett1993} where the state $|\psi\rangle$ of a qubit has to be transferred from the Prover to the Verifier, with the sacrifice of a Bell state $|\beta_{ij}\rangle$. 
The message contains the identifier of the Bell pair to be used in the quantum teleportation process, plus one bit $s$ (selector) to specify whether $|\psi\rangle = |0\rangle$ (when $s=0$) or $|\psi\rangle = (|0\rangle + |1\rangle)/\sqrt{2}$ (when $s=1$) (with uniform probability).
\item The Prover prepares a qubit in the state $|\psi\rangle$ and runs the quantum teleportation protocol, finally sending two classical bits $b_1$ and $b_2$ to the Verifier over a public and authenticated classical channel. 
\item The Verifier completes the quantum teleportation protocol, using $b_1$ and $b_2$.
Then, the Verifier applies $H^s$, then measures the state of the resulting qubit in the computational basis, obtaining a classical bit $v$.
\end{enumerate}  
IF $v=1$ THEN output ``compromised'' ELSE abort.
\end{scriptsize}
\end{bclogo}

\subsubsection{Robustness Analysis of \textsf{AC1}} 
 
\begin{proof}[Proof of Theorem 2]
If the Prover is not controlled by an attacker, it is expected that its behavior is fair, i.e., it does respect the protocol. On the other hand, if the Prover is controlled by an attacker, it could, in theory, send random $b_1$ and $b_2$ to the Verifier. However, this would not affect the chances of being detected by the Verifier. Thus, we assume the Prover respects the protocol also when it is an attacker.

Independently of $|\beta_{ij}\rangle$, if the Prover has not been compromised, the state $|\psi\rangle$ results at the Verifier, after the quantum teleportation protocol has been completed. Thus, by applying $H^s$ to $|\psi\rangle$, the Verifier obtains $|0\rangle$. As a consequence, the successive measurement always result in $v=0$. 

On the other hand, if the Prover has been compromised, the Verifier may get a $|+\rangle$ or $|-\rangle$ state after applying $H^s$. The probability of this scenario is $1/2$, corresponding to
\begin{itemize}
\item $s=1$ and the attacker broke the entanglement by measuring in the computational basis;
\item $s=0$ and the attacker broke the entanglement by measuring in the diagonal basis.
\end{itemize}
In both cases, there is a $1/2$ probability to measure $v=1$, which would reveal that the Prover has been compromised. 
Overall, the probability to detect a non-entangled pair is $1/4$, for each round of the protocol. Thus, the attacker succeeds with probability $3/4$. If the procedure is executed $m$ times, the protocol aborts with probability $(3/4)^m$.
\end{proof}

In other words, \textsf{AC1} succeeds in detecting the attacker with probability $p_m = 1 - (3/4)^m$.  

Importantly, the probability $p$ to detect the attacker in one round of the protocol, remains $> 0$ for any possible choice of the measurement basis by the attacker. 
Assuming that the Verifier gets $|\psi\rangle = \alpha |0\rangle + \beta |1\rangle$, with $\alpha, \beta \in \mathbb{C}$ such that $|\alpha|^2 + |\beta|^2 = 1$, then
\begin{itemize}
\item if $s=0$, then $H^s|\psi\rangle = |\psi\rangle$ and $p = |\beta|^2$;
\item if $s=1$, then $H^s|\psi\rangle = \frac{ (\alpha + \beta)|0\rangle - (\alpha - \beta)|1\rangle }{\sqrt{2}}$ and $p = \frac{|\alpha - \beta|^2}{2}$.
\end{itemize}
Thus, considering that $s=0$ and $s=1$ have the same probability, it is
\begin{equation}
p = \frac{1}{2} \left( |\beta|^2 + \frac{|\alpha - \beta|^2}{2} \right)
\end{equation}

\begin{figure}[h!]
\centering
\includegraphics[width=0.5\textwidth]{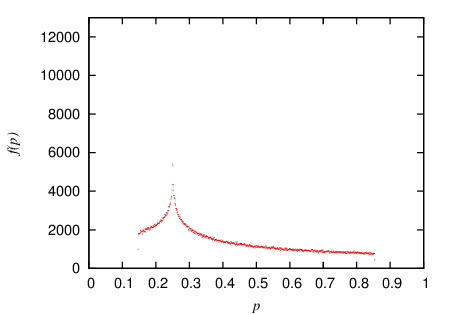}
\caption{Frequency distribution of the probability $p$ to detect the attacker with \textsf{AC1}.}
\label{fig:pDistrAC1}
\end{figure}

To characterize the range of $p$, we generated $10^6$ uniformly random combinations of $\alpha$ and $\beta$ and we observed the frequency distribution $f(p)$ illustrated in Fig. \ref{fig:pDistrAC1}. The domain of $f(p)$ is approximately $[0.146,0.854]$. Interestingly, $f(p)$ has one peak in $p=1/4$, which is the value obtained considering an attacker that performs measurements either in the computational or diagonal basis. Last but not least, the mean value of $p$ is approximately $0.41 > 1/4$.

\subsubsection{Efficiency Analysis of \textsf{AC1}} 

\textsf{AC1} consists of $m$ repetitions of the following operations:
\begin{itemize}
\item classical bit dispatching ($k+1$ by the Verifier, where $k$ is the size of the identifier that specifies which qubit pair is being checked, and $1$ is for $s$; $2$ by the Prover);
\item preparation of a qubit with state $|\psi\rangle$ (once by the Prover)
\item application of the CNOT gate (once by the Prover);
\item application of the $H$ gate (once by the Prover, never or once by the Verifier depending on $s$);
\item qubit measurement (once by the Verifier, twice by the Prover);
\item application of the $X$ gate (never or once, by the Verifier);
\item application of the $Z$ gate (never or once, by the Verifier);
\item binary variable check ($3$ by the Verifier; $k+1$ by the Prover).
\end{itemize} 
The total amount of dispatched classical bits is $k+3$, one more than \textsf{NA2010}.

\textsf{AC1} is more efficient than \textsf{NA2010}, considering the number of sacrificed Bell states. By checking $m=5$ qubit pairs with \textsf{AC1}, the attacker is caught with probability $p_m = 1 - (3/4)^{5} = 0.7626$. With $m=10$, the probability is $p_m = 0.9436$. To get $p_m = 0.99$, it is necessary to check $m=17$ Bell pairs --- with respect to \textsf{NA2010}, $50$\% less Bell states are sacrificed.

\subsection{\textsf{AC2} Protocol} 
\begin{figure}[h!]
\centering
\includegraphics[width=6cm]{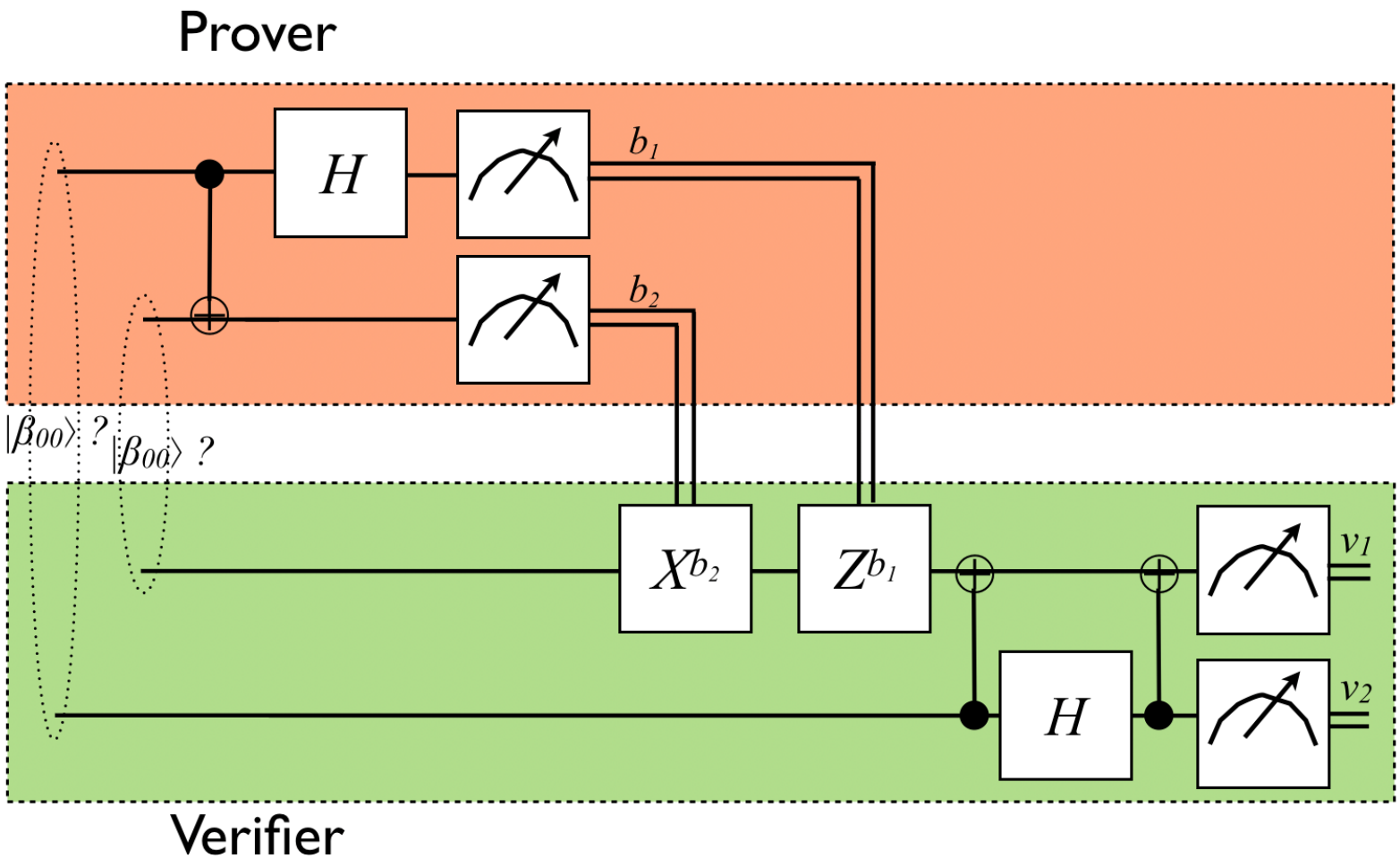}
\caption{Quantum circuit for \textsf{AC2}. The entangled qubit pairs are both supposed to be in the state $|\beta_{ij}\rangle = |\beta_{00}\rangle$.}
\label{fig:AC2}
\end{figure}

To check if the Prover has been compromised, the Verifier may start the \textsf{AC2} protocol illustrated in Fig. \ref{fig:AC2} and detailed in Box \ref{box:AC2}.

\begin{bclogo}[arrondi=0.2, logo=\bccube]{ Box \ref{box:AC2} - \textsf{AC2} Protocol}
\label{box:AC2}
\begin{scriptsize}
\noindent
$v \leftarrow 0$, $k \leftarrow 0$ \quad where $v \in \{0,1\}, k \in \mathbb{N}$\\
WHILE ($(k<m) \wedge (v=0)$) DO
\begin{enumerate}
\item $k \rightarrow k+1$
\item The Verifier selects a qubit pair $(q_1,q_2)$ that is supposed to be in a Bell state $|\beta_{ij}\rangle$. The Verifier owns $q_1$, while the Prover owns $q_2$. Using a public and authenticated classical channel, the Verifier sends a classical message to the Prover, for starting a quantum teleportation process to transfer the state of $q_2$. 
The message contains the identifier of the qubit pair $(q_1,q_2)$ to be checked, as well as the identifier of another qubit pair that is supposed to be entangled and enables the quantum teleportation process.
\item The Prover performs the operations required by the quantum teleportation protocol, finally sending two classical bits $b_1$ and $b_2$ to the Verifier, over a public and authenticated classical channel. 
\item The Verifier completes the quantum teleportation protocol, using $b_1$ and $b_2$.
Once obtained the whole state of the qubit pair to be checked for entanglement, the Verifier applies a quantum circuit whose purpose is to turn the Bell basis into the computational basis. Finally, the Verifier measures both qubits, thus obtaining two classical bits $v_1$ and $v_2$.
\item The Verifier concludes that the Prover has been compromised if and only if
\begin{itemize}
\item the state of the checked qubit pair was supposed to be $|\beta_{00}\rangle$, and $v_1v_2 \neq 00$
\item the state of the checked qubit pair was supposed to be $|\beta_{01}\rangle$, and $v_1v_2 \neq 10$
\item the state of the checked qubit pair was supposed to be $|\beta_{10}\rangle$, and $v_1v_2 \neq 11$
\item the state of the checked qubit pair was supposed to be $|\beta_{11}\rangle$, and $v_1v_2 \neq 01$
\end{itemize}
In that case, the Verifier sets $v$ to $1$.
In any other case, the Verifier cannot decide and leaves $v$ unchanged. 
\end{enumerate}  
IF $v=1$ THEN output ``compromised'' ELSE abort.
\end{scriptsize}
\end{bclogo}

\subsubsection{Robustness Analysis of \textsf{AC2}}  

\begin{proof}[Proof of Theorem 3]  
If the Prover is not controlled by an attacker, it is expected to behave fairly, i.e., to respect the protocol. On the other hand, if the Prover is controlled by an attacker, it could, in theory, send random $b_1$ and $b_2$ to the Verifier. However, this would not affect the chances of being detected by the Verifier. Thus, we assume the Prover respects the protocol also when it is an attacker.

If the Prover has not been compromised, the checked qubit pair results at the Verifier, after the teleportation. The effect of the quantum circuit at the Verifier is to turn the Bell basis $\{|\beta_{00}\rangle, |\beta_{01}\rangle, |\beta_{10}\rangle, |\beta_{11}\rangle\}$ into the computational basis $\{|00\rangle, |10\rangle, |11\rangle, |01\rangle\}$. Thus, if the Prover has not been compromised, a checked $|\beta_{00}\rangle$ always yields $v_1v_2 = 00$ and a checked $|\beta_{11}\rangle$ always yields $v_1v_2 = 01$. Similarly, checking $|\beta_{01}\rangle$ and $|\beta_{10}\rangle$ always yields $v_1v_2 = 10$ and $v_1v_2 = 11$, respectively. 

On the other hand, if the Prover has been compromised, both the teleportation process and the checked qubit pair are affected.
The state of a broken-entanglement qubit pair is then one of $\{|00\rangle, |01\rangle, |10\rangle, |11\rangle\}$ or one of $\{|++\rangle, |+-\rangle, |-+\rangle, |--\rangle\}$, depending on the original entanglement and on the basis used by the attacker to perform the measure. 
For example, let us assume that the qubit pair to be checked is supposed to be in the $|\beta_{00}\rangle$ state and its actual state is $|00\rangle$. The qubit pair indicated by the Verifier for being used in the teleportation process is supposed to be in the $|\beta_{00}\rangle$ state, but actually it may be in the $|00\rangle$, $|11\rangle$, $|++\rangle$ or $|--\rangle$ state. Averaging on all these cases, it turns out that the attacker succeeds ($v_1v_2 = 00$) with probability $3/8$. If the procedure is executed $m$ times, the protocol aborts with probability $(3/8)^{m}$.

The same result is obtained if the state of the qubit pair to be checked, instead of being $|\beta_{00}\rangle$ state, is $|11\rangle$, $|++\rangle$ or $|--\rangle$, and in general for any supposed $|\beta_{ij}\rangle$ state of the qubit pair to be checked and of the qubit pair to be used for the teleportation. 
\end{proof} 

Theorem 3 states that \textsf{AC2} succeeds in detecting the attacker with probability $p_m = 1 - (3/8)^{m}$.  
Further investigation reveals that the probability $p$ to detect the attacker in one round of the protocol, remains $> 0$ for any possible choice of the measurement basis by the attacker. 
Let us assume that the states of the two qubits at the Verifier, after the teleportation, are:
\begin{itemize}
\item $\alpha |0\rangle + \beta |1\rangle$, with $\alpha, \beta \in \mathbb{C}$ such that $|\alpha|^2 + |\beta|^2 = 1$, on the bottom line of the quantum circuit illustrated in Fig. \ref{fig:AC2};
\item $\gamma |0\rangle + \delta |1\rangle$, with $\gamma, \delta \in \mathbb{C}$ such that $|\gamma|^2 + |\delta|^2 = 1$, as the result of the teleportation process.
\end{itemize}

Then, the (CNOT, $H$, CNOT) circuit produces the following state:
\begin{equation}
\begin{split}
|\psi\rangle = \frac{1}{\sqrt{2}}[(\alpha\gamma+\beta\delta)|00\rangle + (\alpha\delta+\beta\gamma)|01\rangle \\ 
+ (\alpha\delta-\beta\gamma)|10\rangle + (\alpha\gamma-\beta\delta)|11\rangle]
\end{split}
\end{equation}

The probability to detect the attacker is
\begin{equation}
p = P\{ |\psi\rangle \neq |00\rangle\} = 1 - \frac{|\alpha\gamma+\beta\delta|^2}{2}
\end{equation}

\begin{figure}[!ht]
\centering
\includegraphics[width=0.5\textwidth]{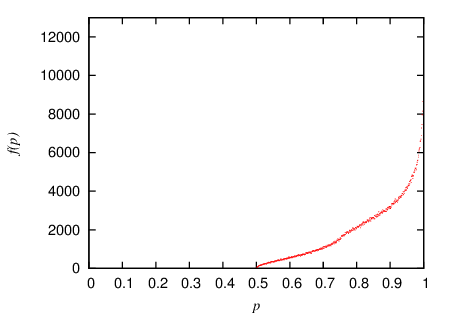}
\caption{Frequency distribution of the probability $p$ to detect the attacker with \textsf{AC2}.}
\label{fig:pDistrAC2}
\end{figure}

To characterize the range of $p$, we generated $10^6$ uniformly random combinations of $\alpha$, $\beta$, $\gamma$ and $\delta$ and we observed the frequency distribution $f(p)$ illustrated in Fig. \ref{fig:pDistrAC2}. The domain of $f(p)$ is $[0.5,1.0]$. Interestingly, $f(p)$ is monotonic with mean value $0.86$, which is larger than $5/8$, the $p$ value obtained considering an attacker that performs measurements either in the computational or diagonal basis.

\subsubsection{Efficiency Analysis of \textsf{AC2}} 

\textsf{AC2} consists of $m$ repetitions of the following operations:
\begin{itemize}
\item classical bit dispatching ($2k$ bits by the Verifier, namely $k$ bits specifying the identifier of the qubit pair to be used for the teleportation, and $k$ bits specifying the identifier of the qubit pair to be checked; $2$ by the Prover);
\item application of the CNOT gate (once by the Prover, twice by the Verifier);
\item application of the $H$ gate (once by the Prover, once by the Verifier);
\item qubit measurement (twice by the Verifier, twice by the Prover);
\item application of the $X$ gate (never or once, by the Verifier);
\item application of the $Z$ gate (never or once, by the Verifier);
\item binary variable check ($2$ by the Verifier, $2k$ by the Prover).
\end{itemize}
The total amount of dispatched classical bits is $2k+2$, i.e., $k-1$ more than \textsf{AC1} and $k$ more than \textsf{NA2010}.

\textsf{AC2} is more efficient than \textsf{AC1}. 
By sacrificing 2 Bell states ($m=1$), with \textsf{AC2} the attacker is caught with probability $p_m = 1 - (3/8)^1 = 5/8$. With the same amount of sacrificed Bell states, \textsf{AC1} allows the Verifier to catch the attacker with probability $p_m = 1 - (3/4)^{2} = 0.43$.
To get $p_m = 0.99$ with \textsf{AC2}, it is necessary to sacrifice $2m=10$ Bell states. With respect to \textsf{NA2010}, $71$\% less Bell states are sacrificed. With respect to \textsf{AC1}, $41$\% less Bell states are sacrificed.

\begin{figure}[!ht]
\centering
\includegraphics[width=0.5\textwidth]{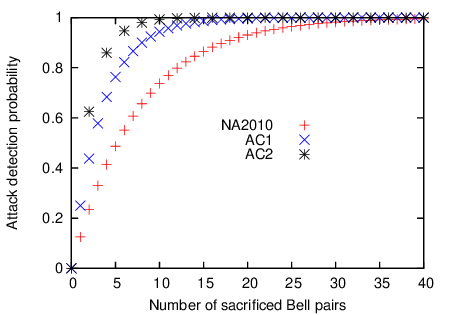}
\caption{Probability of detecting an attacker, with respect to the number of sacrificed Bell states, in \textsf{NA2010}, \textsf{AC1} and \textsf{AC2}. It is worth noting that \textsf{AC2} always sacrifices an even number of Bell states.}
\label{fig:Probabilities}
\end{figure}

In Fig. \ref{fig:Probabilities}, the probabilities of detecting an attacker with the three protocols are compared, with respect to the number of sacrificed Bell states.

\section{Implementation}
\label{sec:Implementation}

We have implemented the aforementioned protocols with SimulaQron \cite{Dahlberg2017}, a novel tool for developing distributed software that runs on real or simulated network end-nodes connected by quantum and classical links. What makes SimulaQron highly interesting is its ability to simulate several quantum processors held by the end-nodes of the network. In this way it is possible to run application software on a simulated quantum network, but also on many inter-connected quantum networks, providing the possibility of simulating a working Quantum Internet. Finally, SimulaQron also provides the feature of allowing software development that is independent from the underlying quantum hardware platform.

\begin{figure}[h!]
\centering
\includegraphics[width=8cm]{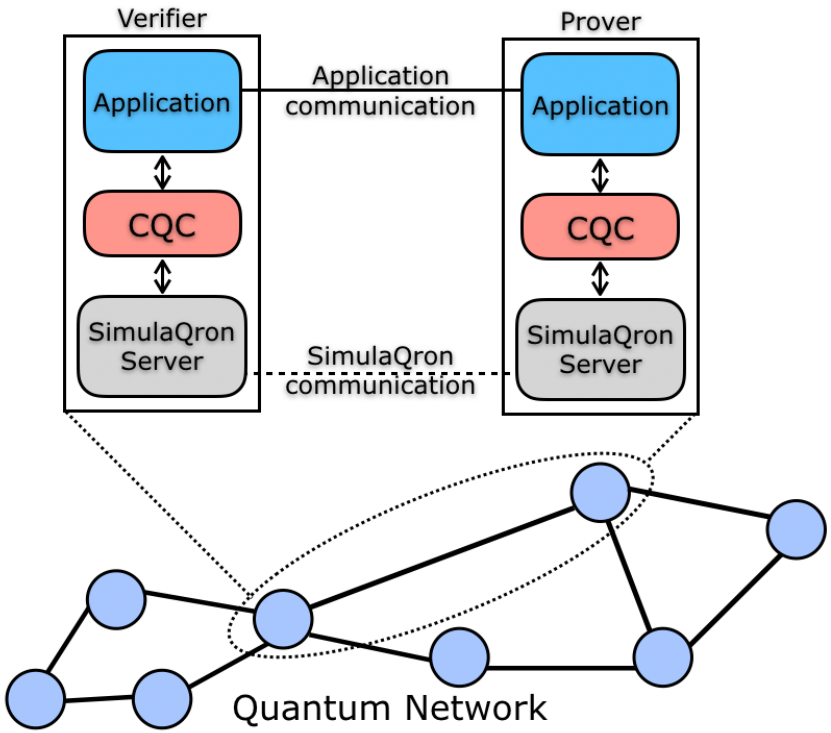}
\caption{Schematic overview of the communication between the simulated nodes.}
\label{fig:simNetwork}
\end{figure}

The Python code we have developed can be freely accessed from the GitHub repository at \cite{EntEvalCode}. For each entanglement verification protocol we have considered in this paper (\textsf{NA2010}, \textsf{AC1} and \textsf{AC2}), we provide the code of the Verifier and the Prover (files \textsf{verifier.py} and \textsf{prover.py}, respectively). In Fig. \ref{fig:simNetwork}, a schematic overview of the communication between the simulated nodes is provided. The simulation of the quantum hardware at each node is handled by an instance of SimulaQron server. As illustrated in the reference paper \cite{Dahlberg2017}, entanglement between distinct nodes is simulated locally by the SimulaQron server associated to one of them. Consistency between nodes is then achieved by appropriate message passing between their SimulaQron servers. The computers in Fig. \ref{fig:simNetwork} can be physically different computers. The CQC (classical-quantum combiner) is a middleware layer that facilitates the implementation of application-level protocols. We emphasize that any node can play the Verifier and the Prover (in Fig. \ref{fig:simNetwork}, two nodes are picked at random).

First, we set $m$, i.e., the parameter that sets the number of sacrificed Bell states (which is $m$ in \textsf{NA2010} and \textsf{AC1}, while it is $2m$ in \textsf{AC2}). We configure the Prover in order that it performs measurements either in the computational or diagonal basis, with equal probability. Then we run the protocol and we count the number of iterations before the attacker is detected (denoted as $m_d$). In case $m_d \leq m$, the attack detection counter $d$ is increased by 1. After $r$ repetitions of the protocol simulation, we estimate the probability of detecting an attacker as $P_m \simeq d/r$.
Simulation results, using $r=1000$, are illustrated in Table \ref{tab:ExpResults} (in brackets, theoretical values are reported). 

\begin{table}[h!]
\caption {Attack detection probability (simulation vs. theory)} \label{tab:ExpResults}
\centering
\tiny
\begin{tabular}{cc|c|c|c|c|c|l}
\cline{3-7}
& & \multicolumn{5}{ c| }{Number of sacrificed Bell states} \\ \cline{3-7}
& & 2 & 4 & 8 & 16 & 32 \\ \cline{1-7}
\multicolumn{1}{ |c  }{\multirow{3}{*}{\hspace{-0.1cm}Protocols\hspace{-0.1cm}} } &
\multicolumn{1}{ |c| }{\textsf{NA2010}} & 0.23 (0.23) & 0.41 (0.41) & 0.66 (0.66) & 0.88 (0.88) & 0.99 (0.99) &   \\ \cline{2-7}
\multicolumn{1}{ |c  }{}                        &
\multicolumn{1}{ |c| }{\textsf{AC1}} & 0.44 (0.44) & 0.69 (0.68) & 0.88 (0.89) & 0.99 (0.99) & 1 (1) &    \\ \cline{2-7}
\multicolumn{1}{ |c  }{}                        &
\multicolumn{1}{ |c| }{\textsf{AC2}} & 0.64 (0.63) & 0.86 (0.86) & 0.98 (0.98) & 1 (1) & 1 (1) &  \\ \cline{1-7}
\end{tabular}
\end{table}

A complete experimental evaluation of the protocols would require a physical testbed involving at least two quantum computers connected by a quantum channel for the initialization step. Currently, there is no publicly available facility. In order to test our protocols on a real hardware, the best we could do was to run the AC1 and AC2 quantum circuits on an IBM Q device \cite{IBMQ}, by means of the Qiskit Terra library. The source code is available in the GitHub repository at \cite{EntEvalCode}. For example, let us consider \textsf{AC1}. Assuming that $|\beta_{00}\rangle$ has been measured in the computational basis, then $v=1$ has probability $1/4$ (as explained in Section \ref{sec:EntVerif}). This is confirmed by the experimental results illustrating the distribution of measured bit configurations, when repeating the protocol execution 8000 times for each possible value of $s$, on the ibmq\_essex device. With $s=0$, the output distribution is: $\{111:3.2\%; 100:1.8\%; 001:37.6\%; 110:3.1\%; 011:3.9\%; 000:44.3\%; 010:4.2\%; 101:1.9\%\}$. With $s=1$, instead: $\{100:9.4\%; 000:15.8\%; 101:16.2\%; 110:12.3\%; 111:9.3\%; 011:12.2\%; 001:11\%; 010:13.8\%\}$. Altogether, $v$ (the leftmost bit) is 1 in about $25\%$ of the measured results.

\section{Conclusion}
\label{sec:Conclusion}

We have illustrated and analyzed three LOCC protocols for entanglement verification across node pairs of a quantum network. Two of these protocols (\textsf{AC1} and \textsf{AC2}) have been proposed for the first time in this work.
Moreover, we have implemented, simulated and experimentally evaluated the aforementioned protocols with SimulaQron \cite{Dahlberg2017} and the IBM Q platform \cite{IBMQ}, obtaining results that confirm the theoretical analysis.

Regarding future work, we will investigate the possibility to extend our protocols to entangled states involving more than two qubits, such as GHZ states, W states and graph states \cite{VanMeter2014}.

\end{document}